# SNOWMASS Calorimetry White Paper

## Novel Low Workfunction Semiconductors for Calorimetry and Detection: High Energy, Dark Matter and Neutrino Phenomena

For the SNOWMASS Committee:
David R Winn
Fairfield University

## Introduction

This White Paper seeks to extend the applications of two weakly bound semiconductor materials, $Cs_3Sb$ and Ag-O-Cs – first developed decades ago for vacuum photodetectors – S1, S-11 photocathodes, respectively - proven to have low electron-hole pair energies. Rather than thin film photocathodes, we propose fabrication of these semiconductors in shapes and volumes which could be used as diode or drifted ion detectors for low energy depositions via atom or few-atomic layers. Ag-O-Cs has an e-hole pair threshold pair low as ~0.5-0.7 eV deposited energy, practical when cooled to $\leq 4°K$; $Cs_3Sb$ – with pair energy as low as 1.6-2 eV - less than Si(3.6 eV), but with thermal noise at room temperature similar to Si at -30°C. Exposure of photomultipliers to 10 MRad doses has shown that thin alkali photocathodes do not degrade more than 2% of initial quantum efficiency.

Progress in atomic layer or few-atom layer deposition includes MBE(molecular beam epitaxy), ALD (Atomic Layer Deposition) pulsed low temperature MOCVD, and precision e-beam evaporation. Over the past ~decade these techniques been shown to deliver large area (~$m^2$) and films of up to few mm thicknesses. These techniques offer the potential to fabricate multi-element near crystalline semiconductor performance than has been practical only with thermally grown bulk crystals, heretofore.

Such detector tiles could be fabricated by atomic or few-atomic layers on atomically flat substrates, prepared before semiconductor deposition with ~10 nm thermal and stress grading films and a contact back electrode. The bottom electrode may be Ohmic or form n-p-, or Schottky- diodes. After an atomic or few-atom deposition of the semiconductor, a top electrode(Ohmic or Diode) is deposited.

Recently single layer BN films or graphene can both exclude oxygen and water vapor and yet remain transparent to photoelectrons. A single layer of graphene alone excludes even He and forms a top electrode and a nearly transparent seal to low energy radiation ($\alpha$, $\beta$, $\gamma$). This technique can also be applied to hydroscopic scintillators such as CsI or NaI. We propose that a research program be initiated in academic and DOE labs with nanofabrication and test facilities. Completed novel semiconductor detectors should be tested for mobility, carrier lifetime, I-V, sensitivity to radiation and radiation damage. Atomic or few-atom heterogenous layer-by-layer assembly offers the potential to make 100nm~mm stacks of alternating materials to form multi-atomic semiconductors. Atomic layer assembly is capable of incorporating dopants for diodes, electrodes, transparent electrodes, hermetic seals, and materials serving as neutron converters.

**Applications** of semiconductor detectors in elementary particle physics include detection of low pair-energy thresholds range over detecting dark matter interactions, low energy (sub-MeV) neutrino interactions, neutrino-less double-$\beta$ decay, and neutrino interactions from the cosmos or sun may result. The 2 materials have been shown to be radiation resistant as tested in photomultipliers, and with mobilities of both e and holes 2-4x higher than Si, could be used for high rate, high resolution sampling calorimeter sensors similar to Si sampling calorimeters or tracking(more ion pairs per dE/dx, higher dE/dx and lower Xo).

## Physics Cases

In high energy, intensity and astroparticle physics frontiers, signals of energy, time and position via signals from energy deposits from photons, neutrinos and charged and neutral massive particles has lead to major discoveries. Better signal resolution, large signals per energy deposited, and very low energy detection thresholds are of great interest for dark matter searches, one of the outstanding issues is fundamental physics. The energy deposited in a detector by many possible interacting dark matter particles may be as low as or even lower than ~10's of eV. Event detection at such energy levels requires experimental apparatus at very low temperatures to distinguish the deposited energy from the thermal energy in the detectors. For example, the threshold energy deposition for LUX and XENON 1T are at best ~1-2 KeV[1]. SuperCDMS uses Si and Ge with pair energies of 3.6 and 2.98 respectively together with phonon detection and an energy threshold of 70 eV[2] at mK° with the phonon signals.

If an effective e-hole pair energy were 0.7 eV-0.4 or lower (example: Ag-O-Cs semiconductors) and ~100 electrons could be collected from a deposition of ~0.1 KeV, a ~10% energy resolution in principle would result. If a sub-eV diode or drifted ion detector energy detection is possible, low energy neutrino interactions could be studied, including solar/reactor/radiosource (anti)ν oscillations, ν coherent interactions (at present not well studied - the neutral current ν-scattering "floor" for DM searches from solar and cosmic ν's), or even ν-mass experiments and ν-less double-β decays. Energy waveform measurements (>100MHz sampling), integrated energy depositions in at high rates (≥100MHz at future colliders and tagged neutrino beams) and radiation resistance are important for calorimetry, tracking and fast timing in many existing and future experiments in the energy & intensity frontiers.

A semiconductor with a faster signal, with a larger signal per path length, lower thermal noise and higher radiation resistance compared to Si tiles would be advantageous). We propose repurposing and re-engineering S-1(Ag-O-Cs) and S-11($Cs_3Sb$) - weakly bound semiconductors used chiefly as photocathodes, heretofore. A key challenge for semiconductor ion (e-hole) detectors which the 2 materials address:
   - 0.5-2 eV energy detection threshold and resolution.
   - Semiconductor detectors also benefit from high atomic number (Z), high density (ρ), low electron-hole pair production energy $E_{pair}$ or $E_p$ (also called $E_{threshold}$ for e-hole pair production) – see Table 1 below.

In some applications in the energy and intensity frontiers with calorimetry and ionizing particle detection, problems include rate and timing (LArgon), radiation damage (scintillators, Si), and rates exceeding 100's MHz(tagged neutrino beams), as issues for ionization detectors. The mobility of the e-holes in S-1, S-11 cathodes are approximately equal (unlike Si or GaAs), and very large compared to Si or GaAs (see Table 1 below), meaning that polarization at high rates is lower, and the potential for 10's psec time resolution may be better than Si detectors in calorimetry or timing. Experimental evidence for radiation resistance from irradiating photomultipliers at high MRad levels has shown that radiation damage to vacuum photocathodes is negligible [3]- PMT damage is due entirely to glass envelope glow and darkening.

|        | Z     | ρ(g/cc) | Egap(eV) | $E_{pair}$ | $\mu_e$, $\mu_{hole}$ cm$^2$/V/s | Lrad(cm) |
|--------|-------|---------|----------|------------|----------------------------------|----------|
| Ge     | 32    | 5.3     | 0.7      | 2.98       | ≤3900, ≤1900                     | 2.3      |
| Si     | 14    | 2.3     | 1.1      | 3.6        | ≤1400, ≤450                      | 9.4      |
| GaAs   | 31,33 | 5.3     | 1.4      | 4.4        | ≤8500, ≤400                      | 2.3      |
| *$Cs_3Sb$* | 55,51 | 4.6 | 1.6      | 2.0        | 10,000-10$^6$                    | 1.9      |
| *Ag-O-Cs* | 55,47,8 | 7.1 | <0.3    | 0.4-0.7    | Predicted ~5,000                 | 1.8-2.0  |

**Table 1: Semiconductor properties for low energy deposition detectors.**

*Jiffy Diode Ion Detectors:* For the lowest bandgaps $E_g$, operation must be cooled to LHe or lower to avoid thermal noise generation. The probability per unit time that an electron-hole pair is thermally generated is $P(T) = CT^{3/2} e^{-E_g/2kT}$ where T is absolute temperature, $E_g$ is the band gap energy, k is the Boltzmann constant and C is a proportionality constant characteristic of the material. For operation at room temperatures, a bandgap Eg greater than about 1.3 eV is necessary so that thermally generated carriers do not dominate detection of low energy events. For example, Si (Egap ~ 1.1 eV) often requires cooling; SiPM(Si Photomultiplier) often require cooling to -30C° for sufficiently low noise single p.e.(Photoelectron) detection. For good resolution, the pair energy Ep is given by Ep=Eg +Ea (Ea=electron affinity), also called by some authors the threshold energy $E_{th}$. $E_{th}$(=Ep) is similar to/related to the work function $\phi$ but can be more or less than the vacuum level VL. Fig. 1 shows generic energy bands levels for semiconductors divided into 3 classes. The levels are relative – that is $E_{th}$(or Ep), the valence band VB and the conduction band CB are all shown as equal energies in the 3 classes, in the same proportions or $\Delta E$ to each other (not always the case for any particular semiconductor). The left and middle describe semiconductors where the threshold energy for pair production is above the vacuum level VL. In that case, when e-hole pairs are produced, the electron has a chance to escape, as its energy can be above the vacuum level, and can be a photoelectron (p.e.) or photoconducted electron. The best semiconductor material for photoemission (photocathodes – PC) have energy levels qualitatively similar to the left figure; once a pair is made, over a large range of energy depositions, the electron can escape, while the middle diagram describes photoemitters with a restricted range of emissions. Multi-alkali and GaAs photocathodes are like the left most diagram, while most mono-alkali PC are like the middle diagram.

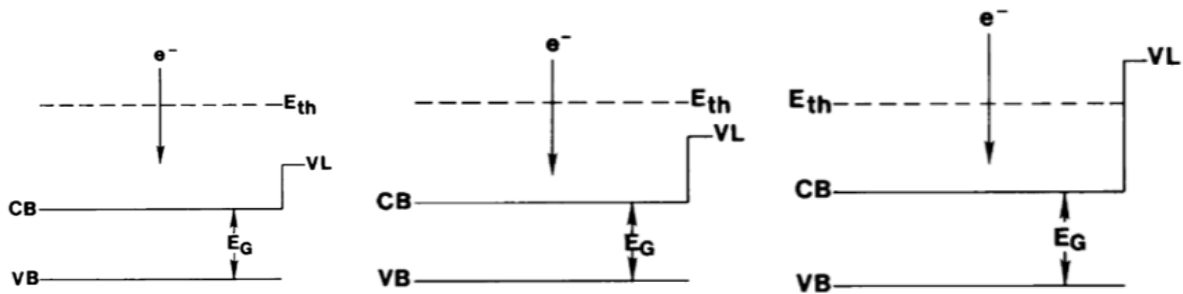

*Figs 1:* Three ideations of classes of semiconductors: VB=Valence Band, CB=Conduction Band, VL=Vacuum Level, $E_G$=Band Gap, $E_{th}$ is Ep, the threshold energy to produce electron-hole pairs.

In alkali photocathodes typically the work-function is slightly more than Eg+Ea (Fig 2). Poor photoemitters are like the right diagram in Fig 1 - like Si or Ge - where most of the electron-hole pairs do not have enough energy to be above the vacuum level (work function) to escape. However, the electron-hole pairs are above the conduction band and can be drifted to electrodes or a junction.

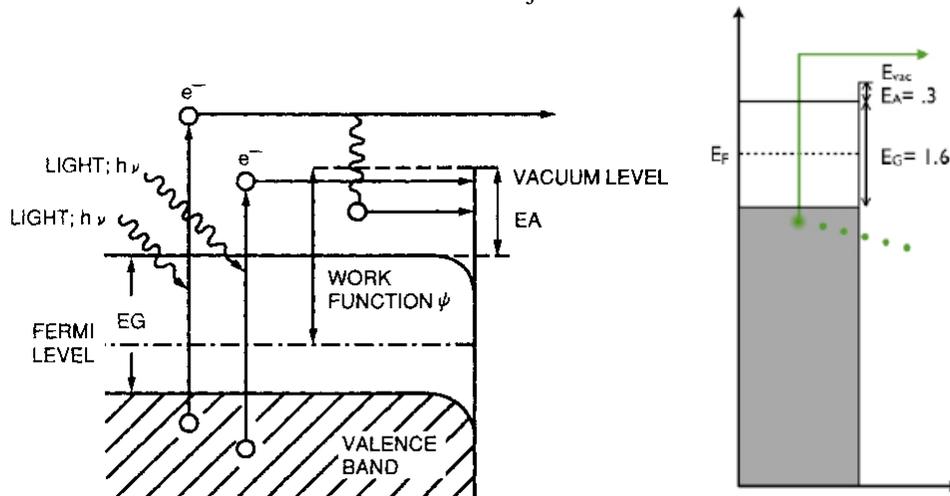

*Figs. 2:* Left: Energetics for alkali photocathode vacuum emission – the pair energy Ea+Eg is near the vacuum level. Right: $Cs_3Sb$ semi-conductor energy level band diagram – the energy for pair production is ~Ea+Eg= 1.9-2.0 eV, near the vacuum level and like the 2nd class shown in the middle of Fig.1.

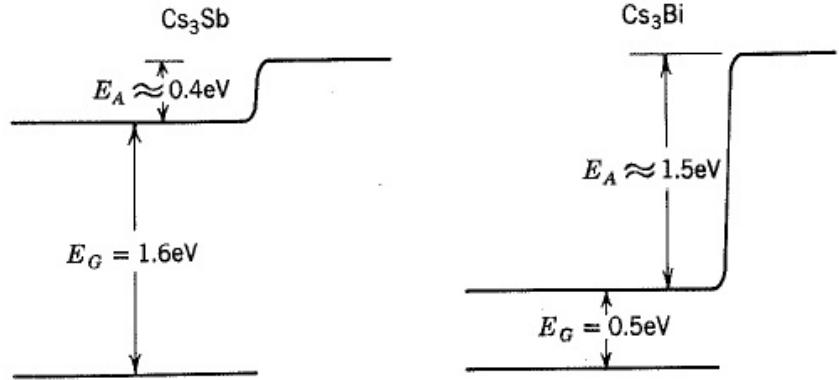

***Figs. 3:*** *$Cs_3Sb$ and $Cs_3Bi$ have nearly equal pair energies* $E_p$ proportional to $E_A+E_G$ ~2 eV, *but very different Eg*. Si is analogous to $Cs_3Bi$, with a low $E_G$ so that thermal carriers more easily generated and a source of detector noise. $Cs_3Sb$'s large $E_G$ suppresses thermal carriers/noise relative to Si, but has a lower $E_{pair}/E_A$.

Figs. 3 show that $Cs_3Sb$ and $Cs_3Bi$ have nearly equal pair energies $E_p$ proportional to $E_a+E_g$ ~2 eV. However, $Cs_3Sb$ has a bandgap 1.6 eV, considerably larger than the 0.5 eV of $Cs_3Bi$. That large $E_g$ inhibits thermal energy from promoting carriers above the Fermi level, and therefore can operate with minimal or no cooling for detecting many types of low energy events, which could include very low energy neutrino scatterings. As an example, *Si has a similar $E_a/E_g >1$ as has $Cs_3Bi$ – Si has a low $E_g = 1.1$ eV and therefore requires cooling for many applications such as SiPM or Si diode detectors compared, with $Cs_3Sb$ with 1.6 eV Eg*. However, Si has an $E_p=3.6$ eV compared with $Cs_3Sb$ $E_p=2$ eV – so $Cs_3Sb$ can operate at room temperature with more signal than Si.

For a given energy deposition in a detector material (the full energy photoelectric peak), to zeroth order the energy resolution limit $\sigma$ is given by $\sigma \sim \sqrt{(F/n)}$, where $n \sim E/E_p$ is the number of generated carriers, and F is the Fano factor which may be as small as ~0.1 for semiconductor detectors (Ge~0.06). A basic figure of merit M for a detector material (independent of geometry) at energies <500 KeV is roughly $M \sim \rho Z^{3.5}/\sigma$. For higher energy, the reciprocal of the radiation length $L_{rad}$ which scales as a lower power of Z ie $\sim Z^2$.

A key issue in multi-atomic semiconductors is defects from impurities and lattice dislocation, as sources of noise and of traps inhibiting carrier lifetimes/lengths. Semiconductor fabrication of GaN, hex-BN, SiC InGaN, ZnO:X, AlGaN have had success in suppressing defects to ~10-1,000/cm² in thin (up to ~100 µm thick) layers using industrial PVD(physical vapor deposition) and MOCVD(metal-organic CVD) processes. We propose exploring atomic or few-atom layer techniques for production of near perfect thin diodes. If 50-100µm thick tiles as rad-hard detectors are possible - like the Si used in trackers or in the CMS upgrade forward region sampling calorimeter (100-200 µm thick si tiles).

If only few nm thick tiles can be defect free, the semiconductor can be oriented so charge is drifted parallel to the tile plane, similar in concept to vertical FETs. In principle, ≤ 100 molecule thick defect free semiconductor films in ~cm² areas, atomically assembled on lift-off BN or graphene (see Fig 4 below) or similar, can be stacked together, turned 90°, with the electrodes on opposite tile edges, and then read-out drifting parallel to the film planes to form larger detector volumes.

There are many examples of semiconductors as diodes or drift cells for calorimetry (Ge, CdTe etc.). Ge is a superior semiconductor detector, reaching resolutions of ~0.2% at 662 KeV and peak/Compton ratios of 30 or even higher, with $E_g = 0.7$ eV and $E_p = 2.96$ eV; *but too large for extending present Dark matter searches*. Silicon pair energy $E_p=3.6$ eV is similarly too large, as is GaAs. By contrast, semiconductor materials normally used for vacuum photocathodes have much lower $E_p$. $Cs_3Sb$ (S-11 photocathode-like) has a ~2eV pair energy. Cs-Ag-O (S-1 like) has an averaged pair energy/work function Ep=0.7 eV in commercial applications [4],[5], but studies have shown that in small patches as fabricated in research studies that the pair energy/work function is a remarkable Ep = 0.4 eV [6].

We propose using atomic layer assembly techniques [ALD -atomic Layer Deposition; MBE – Molecular Beam Epitaxy; Flood MBE; low-temp pulsed-CVD – Chemical Vapor Deposition and variants] to make precisely structured Ag-Cs-O – in the hope that the lowest pair energy can be achieved, and $Cs_3Sb$ for sensors with low energy threshold operating at room temperatures. Atomic/few-atom layer assembly techniques can be used to protect the cesiated materials from air and water vapor with films that are effectively transparent to radiation – single layer graphene is now a standard deposited material; 1 layer excludes He, oxygen and water vapor (on CsI, NaI?).

In vacuum photocathodes the photoelectron escape depth is at most ~100's of atoms thick, with almost no electric field enhancement from the ~100-300V between cathode and nearest electrode or dynode. In the application proposed here, a biased drift device or diode junction device would not have that restriction in thickness with an applied field, similarly to GaAs based photocathodes vs GaAs diodes to collect the electron-holes. It is possible - depending on the metal electrodes -that a Schottky diode could result with the p-type $Cs_3Sb$ or Cs-Ag-O, or a few nm thick defect or low-doped layer could be deposited to form an n-p junction.

*Hermetic Issues:* Cesiated materials and many ionic-compound scintillators must be protected from water and oxygen. Recently, practical deposition of single atomic layers of boron nitride (BN) and graphene on $Cs_3Sb$ have been shown to exclude water and oxygen, yet remain transparent to photoelectrons – an extraordinary possibility with wider applications.

**Photocathode Bulk Semiconductor Pilot Projects:** We propose Pilot Projects to study the potential of 2 atom-layer or few-atom-layer-assembled semiconductor materials as e-hole detectors - used heretofore as diffused gradient amorphous thin films for vacuum photocathodes (S-1, S-11), with e-hole pair energies (0.4-0.7) eV and 2.0 eV respectively. The goal of such projects are to study the potential methods such fast MBE, flood MBE, plasma enhanced MOCVD, ALD, pulsed MOCVD, and laser or e-beam controlled evaporation to fabricate near crystalline semiconductor tiles and measure performance as detectors, that could be scaled up for potential applications in fundamental physics experiments.

*1. Cesium antimonide ($Cs_3Sb$)* is a weakly bound cubic semiconductor, lattice constant of 9.15 Angstroms, normally p-type from Sb excess displacement defects, with relatively well-known properties[7]: a very low electron affinity of 0.4-0.5 eV, and room-temperature intrinsic resistivity ~1,000 Ω-m, about $10^{7-8}$ times higher than Sb, comparable to the 500 Ω-m of Ge at 77K, and a mobility μ - even in the highly defect filled films of thermal diffusion fabricated photocathodes - exceeding $10^4$ $cm^2$/Vs. Indeed, on theoretical grounds, μ could exceed $10^6$, like InSb[8]. With precision assembly of films or tiles using Atomic/few atom Layer Fabrication, displacement defects may be minimized, and may become more intrinsic rather than p-type. With these attributes, the polarizing of the material due to differing electron and hole mobilities, together with the overall high mobility, imply that very high rates of ionizing events may be sustained, exceeding 300 MHz, and ~10 psec timing is feasible.

Recently it has been shown that one molecular thickness of boron carbide (BN) or one layer of graphene can protect $Cs_3Sb$ from air, important for practical detectors with minimal effects from air. Remarkably, the BN layer *lowers* the pair energy from 2 eV to 1.6 eV as shown in Figures 4, 5 [9], [10] whereas graphene layers raise the pair energy[11]. BN and graphene atomic layers deposit at T<100°C [12].

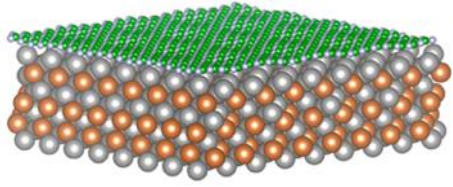

*Fig 4:* Molecular model of layered Cs$_3$Sb with a BN (boron nitride; graphene similar) single molecular layer top-side covering. The e-hole pairs could be drifted parallel to the atomic planes in perfect crystals that result from (near)atomic layer assembly.

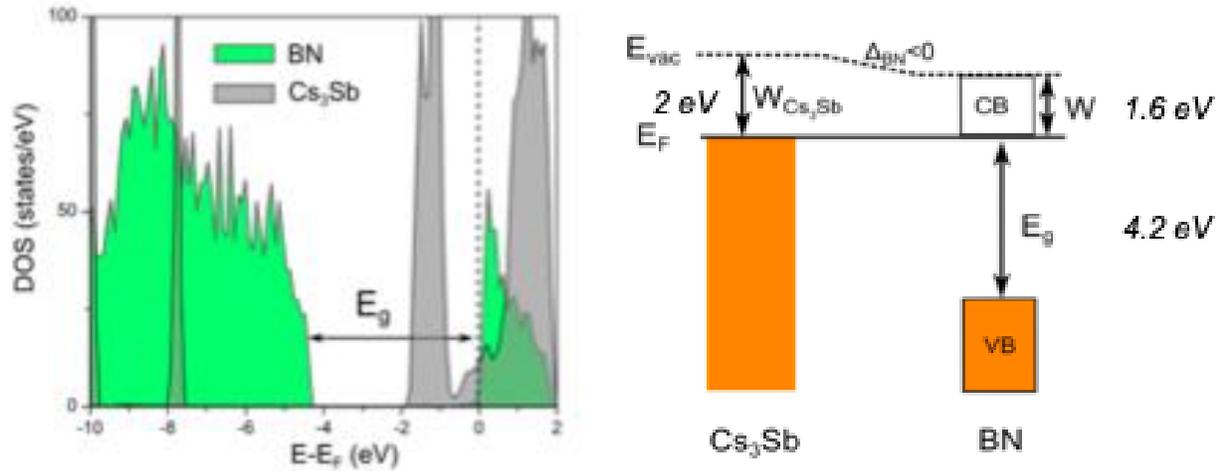

*Figs 5(L):* Energy bands of Cs$_3$Sb+BN. *(R):* Remarkably, the implied electrostatic potential *lowers* the work function of Cs$_3$Sb from 2 eV to 1.6 eV.[9],[10], [11]

For calorimetry, as compared with Ge at 2.98 eV, Cs$_3$Sb pair energy is a remarkably low 2 eV [13], [14], [15] or 1.6 eV if coated with BN as discussed above, and therefore, in principle, Cs$_3$Sb has an intrinsic resolution better than Ge by ~$\sqrt{(2.98/2)}$ ~1.22 or $\sqrt{(2.98/1.6)}$ ~1.36. On the other hand, the bandgap is 1.6 eV, exceeding Si (1.1 eV) and even CdTe (1.47 eV) so that thermal noise is minimal at room temperature. The thermal noise current at room temperature is between 10,000-20,000 times lower in Cs$_3$Sb than Silicon, which alone makes it interesting as a detector, *if defects and impurities are controlled by atomic layer fab*. That the pair energy Ep is only 0.4 eV above Eg is the secret to this potentially breakthrough material, quite unlike Si, Ge and the like. The density of Cs$_3$Sb is 4.6, compared to 5.3 for Ge, however Z of Cs$_3$Sb is 55 and 51 (average ~54), compared with 32 for Ge. The result is that photoelectric absorption – as an example of a radiation detector function - is about 4-5 times higher for Cs$_3$Sb compared with Ge for any given thickness at ~662KeV gamma-rays, and the Peak/Compton ratio should exceed Ge, provided a low-defect crystal can be obtained. Moreover, the radiation length is short.

As a detector for low energy deposition photons or dE/dx, Cs$_3$Sb may be superior Ge in many respects. Because of its larger atomic number its photoelectric absorption is larger, its peak to Compton ratio is larger, and its radiation length is shorter. It has better resolution because its pair energy is lower. It does not need to be cooled for applications at room temperature, because its bandgap is large, even larger than for Si. Compared to Ge, the cost of the highly purified raw material is negligible, albeit the fabrication costs are at present high. The speed of a Cs$_3$Sb detector could be 5-10x faster than Si, since the mobility is substantially larger, by factors of 7-700, with nearly equal e and hole mobility, making it able to take high rates without polarizing. A shortcoming is that, like NaI, it must be protected from the atmosphere. It can be grown from a melt almost identically to that of NaI or CsI fabs, which may be justified in few mm thicknesses. Crystals of Cs$_3$Sb have been grown readily by an inert atmosphere or vacuum thermal melt, heating Sb powder with Cs metal particles at 725°C – Figure 6 - and their bulk properties measured [16], but we know of no studies of Cs$_3$Sb used as a dE/dx detector. The technology to grow crystals of CsI and NaI occur at similar forming temperatures and such commercial facilities could be readily adapted to Cesium Antimonide. Today we can obtain Cs and Sb with much higher purities than the earlier experiments, circa ~60 years ago. Because of the possible benefits in many applications (and cost-savings) the cost and effort to grow bulk materials and test them as low energy deposition detectors is an interesting addition to very low energy calorimeters, but would be addressed after success of this Pilot Proposal.

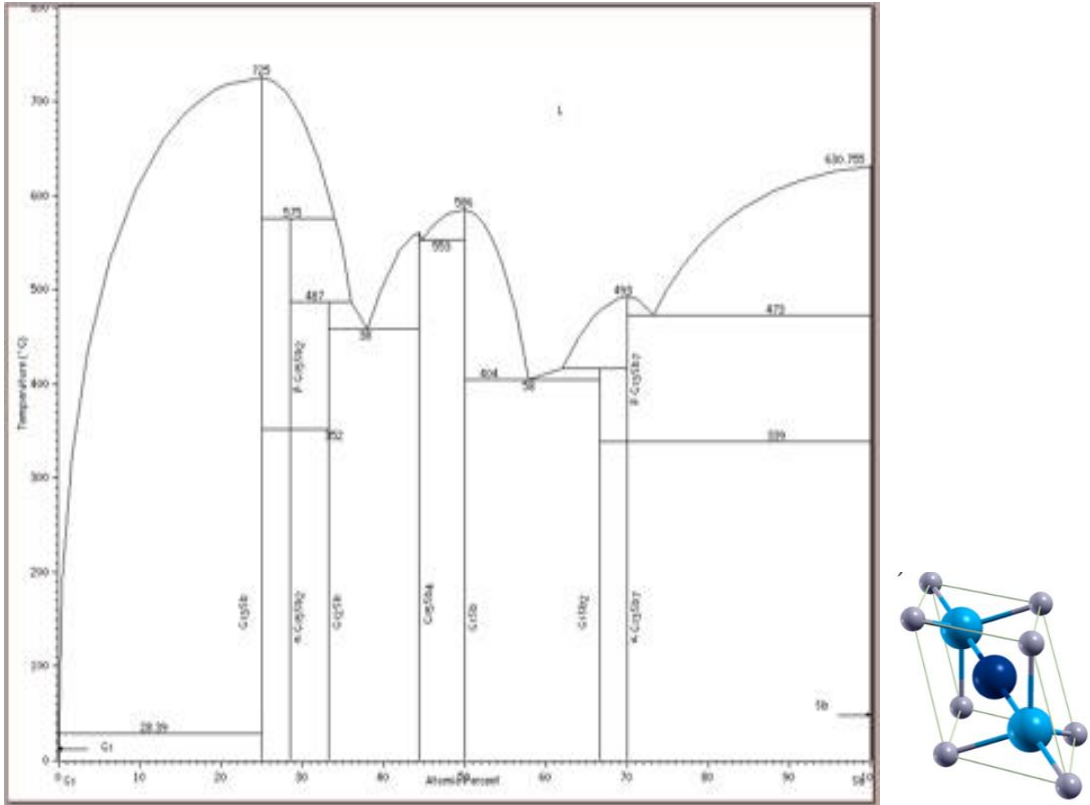

**Figs 6:** - Phase diagram for Cs$_3$Sb melts – Temperature vs atom% of Sb - perfect 25% stochiometry forms at 725 °C (the left-most peak). R: Cs$_3$Sb peroskovite– note expansion of Sb(grey) due to 3 Cs(blue).

**2. Ag-O-Cs** is the basis for the S-1 photocathode. It is hardly used any more for vacuum photoelectron detectors since it has a noise/dark current at room temperature 4-6 orders of magnitude greater than the best bi-alkali photocathodes and 3-4 orders of magnitude larger than the Cs$_3$Sb S-11 as shown in Fig.9. We propose turning that bug into a feature. Ag-O-Cs has the lowest work-function or pair energy of any other material known, as low as 0.4 eV [17], [18]. A few mm thick tile cooled to <1K° could be a superior Dark Matter Detector, capable of generating ~100 electrons for ~50 eV deposited. Deposition on Si may lower it further by band bending or by the electrostatic potential induced by protecting it with an atomic layer of BN and perhaps by graphene on top of the BN. Fig. 7 shows the structure of Ag$_2$O amenable to incorporation of Cs. The basic reaction forming an Ag-O-Cs cathode is represented by

$$Ag_2O + 2\ Cs \rightarrow 2\ Ag + Cs_2O.$$

The silver atoms remain in the lattice and are essential for the energetics. Fig 8 shows optimal Cs:O ratio of 1:1. The process proceeds at modest temperatures 100 C°-300 C°. The standard formation processes of Ag-O-Cs photocathodes are described in detail in A.Sommer's monograph on photocathode formation[19], where the silver atoms as part of the material make it difficult to be transparent to visible light. We emphasize that in a diode/ion chamber, there is no need to preserve semi-transparency, and the Ag content can be optimized for semiconducting properties rather than semi-transparency and photoemission.

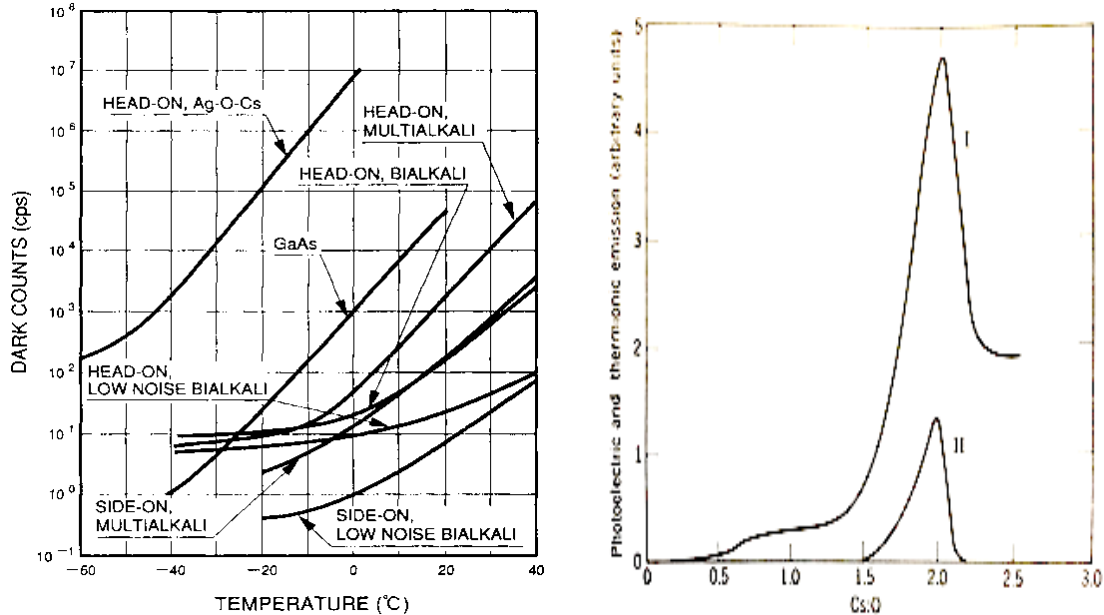

***Fig.7(Left)***. Photocathode Dark Currents in PMT photocathodes[Hamamatsu[20]] vs T. *The Ag-O-Cs dark rate is ~4 orders of magnitude larger than others from the very low energy needed to generate carriers.*
***Fig 8:*** (Right) Electron emission (arb units) vs Cs:O fraction during reaction of silver oxide with Cs; photo electron emission(curve I) thermionic emission (curve II) [21] – *the material is optimized when the Cs:O ratio is 2.0.*

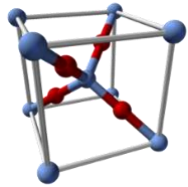

***Fig 9:*** a: The structure of $Ag_2O$ is amenable to atomic layer assembly; b: incorporating atoms in layers of Cs, despite a large Cs atomic radius, forms the Ag-O-Cs semiconductor

**Key issues** which will determine $Cs_3Sb$ and Ag-O-Cs feasibility include finding answers to the following dozen questions:
  (1) What is the surface micro/nano crystallinity and composition viewed in an SEM equipped with 2 of XAFS(X-ray absorption fine structure), XPS(x-ray photoelectron spectroscopy), XRD (X-ray diffraction) and EDX(energy dispersive x-ray spectroscopy) equipment.
  (2) Can Al, Ni, Cr, NiCr electrodes be Ohmically or Schottky contacted?
  (3) Is it n-/p- type/intrinsic, and can excess/deficit of c?
  (4) Can excess or deficit of one of the constituents make p- or n-type films?.
  (5) What is the effective resistivity of a bulk crystal, and as a function of temperature?
  (6) What is the shape of the I-V curve?
  (7) What are the carrier mobilities $\pm\mu$, lifetimes, pathlengths, and effective depletion depth?
  (8) What is the photoconductivity as a function of wavelength?
  (9) What is the thermal current noise as functions of T?
  (10) What is the breakdown voltage and electric field?
  (11) What is an adequate annealing time?
  (12) What is the response to ~5 MeV alpha particle, $^{56}Fe$ x-rays, and muon or electrons?

(13) …..

*Outline of possible fabrication of Semiconductor Diodes/Ion Chamber Radiation Sensors:*
For initial feasibility and control, we propose starting with small 2"- 4" atomically flat wafer substrates: either $SiO_2$, alumina ($Al_2O_3$) or Si wafers (pure, n-, p- doped, or compensated doping)- and with or without $SiO_2$ 10-20 atom thick insulator films and at least 1 stress-matching film. The substrates could be coated with a conducting metal film as a bottom electrode, typically nichrome, Al or Au or matching amalgams (ex:Ge-Au), covering the entire wafer, with a removable resist forming a removable ring of at least 2-3 mm wide on the outside diameter, for the bottom electrode connection.

Ohmic, diode, and/or Schottky contacts are necessary, and will be of particular study. If necessary for adhesion or thermal expansion, Alumina substrates may be prefered, with an intermediate stress relief film of ~1-2 nm. The interior of the wafer ticket -inside the resist ring- is the semiconducting material, deposited by the proposed atomic or few atom layer processing. On completion of the semiconductor, either an Ohmic conducting or a metal Schottky diode top electrode is fabricated.

Vacuum protective layers of the cesiated material is necessary, as these are destroyed by oxygen or water vapor. A single layer of graphene excludes all gasses - even He - yet remains conductive; a single or double layer BN (especially for the cesium antimonide to lower the pair energy) has acceptably low resistance See Figs 4,5).

A thin film top metal electrode would be over-deposited. These fragile top films would be protected from mechanical damage by a thin, strippable high vacuum/low out gas NASA qualified polymer (polypyroles or others), and the sides of the ticket would be protected from air by deposition of a ring of the height of the semiconductor, and width of ~2mm, protecting the sides of the atomic layered semiconductor. The outer ring is stripped off exposing the bottom connection.

The compound semiconductor tiles on the ticket would be characterized immediately after fabrication using some of the XRD, XPS, XAFS, EDX systems auxiliary in SEM's, and with AFM before being sealed. Electronic measurements of e and hole mobility, resistivity, I-V, carrier lifetime, and response to light and ionizing radiation will be carried out n using an SEM with additional probes and vacuum feedthrus.

*Deposition of Crystalline $Cs_3Sb$*
MBE: Molecular Beam Epitaxy is the most sophisticated form of PVD (physical vapor deposition) - capable of monolayer deposition - is proposed as the baseline process– a compromise between ALD and CVD. MBE is not very commercial because of the constraints of very high vacuum make large scale use prohibitive, but it can produce pristine films of Sb at 1 monolayer per second or 1μm per hour, much faster than most ALD. A 1nm Sb film when cesiated becomes 7nm thick. It is reasonable to expect 2-3μm per hour thicknesses with alternating MBE of Sb and Cs. Present MBE rigs use Sb for producing antimonide GaSb and AlSb films. An MBE deposition port for Al can be readily adapted to Cs with a standard precision UHV valved alkali metal source.

Insulating atomically flat substrates are first coated with a ~10-20 atom layer deposition of Ni, Al or preferably nichrome as a back electrode and potential Schottky diode contact. An uncoated Si(100) substrate is a pseudo-lattice match to $Cs_3Sb$, and if n-doped possibly could form a diode with p-type $Cs_3Sb$. Collimated Cs vapor will be pulsed after each 1-2 Sb layers, introduced as sufficient for $Cs_3$ layers. This techniques is far more ideal than the usual way of photocathode fabrication, where Sb is thermally deposited to a desired thickness (20-25nm), and then Cs vapor diffuses into the Sb - there is a gradient through the thin film - this cannot create a perfect material, in contrast to the layer-layer build up proposed here. Tiles could be ALD-coated with BN to both protect it and lower the pair energy to 1.6 eV, and one with graphene as a test of protection from air with interesting potential for a vanishingly thin to radiation Ohmic contact for low energy particles.

Several assemblies of 2"-4" major dimension sizes on different prepared substrates could be exposed simultaneously. A run of 10 days, including over long weekend, could produce a film 0.5-0.75 mm thick, enough for demonstrating the properties needed for a pilot project measuring properties indicative of continuing to larger radiation detectors. If scheduling is restricted to 1-2 days at a time, and multiple exposures are needed for full thickness, a vacuum load lock stores tickets between exposures, and/or to be able to transfer tickets between equipment for further processing (see the Cs-Ag-O task below), in particular to produce graphene and BN films on the top surface and sides of the $Cs_3Sb$ and top electrodes of AL, Ag, Nichrome, and possibly other after investigation of adhesion and thermal expansion.

ALD: An alternative is ALD (atomic layer deposition) [22] of antimony and atomic layer PVD (physical vapor deposition), now standard in VLSI fabs. ALD is the best method to create pure conformal films; elemental antimony can be formed by ALD using $SbCl_3$ and $(Et_3Si)_3Sb$ as precursors. ALD of a few atomic layers will be used to prepare some of the substrates.

CVD: Temperature tuned CVD (chemical vapor deposition) is faster than ALD, and can be nearly as good, but does not create as pristine nstochometric-perfect and atomic-layered materials. Single precursors include tris (trifluoromethyl)stibine, $Sb(CF_3)_3$, Lewis base adducts of $Sb(CF_3)_3$, and antimony trihydride(stibine). Cyclic-pulsed plasma-enhanced chemical vapor deposition using $Sb(i-C_3H_7)_3$ and $H_2$ plasma is faster and may prove competitive at lower T ~275-350°C. Control of the deposition temperature varies the film properties, such as resistivity, surface roughness, and crystallinity. The film growth rate (thickness/cycle) is ~0.10–0.5 nm/cycle[23]. Pure antimony films can be deposited at temperatures below 300°C with growth rates approaching 20nm/min using a low pressure hot-wall CVD reactor[24]. The film growth rate in pulsed mode (thickness/cycle) is ~10–0.5 nm/cycle. [Note: ALD and pulsed CVD Tools have been developed to coat very large areas (~2m$^2$) for advanced image displays(TV/video).]

## *Deposition of Ag-O-Cs*

*Ag-O-Cs:* To make the S-1-like material, Cs is applied by PVD alternating with 2 atomic layer depositions (ALD, Pulsed CVD, or plasma enhanced PVD) of silver oxide $Ag_2O$. The deposition will be on pseudo lattice match to both n-type and p-type Si electrodes with an atomically flat 4" wafers, to determine if the Ag-O-Cs turns out to be p- or n- type, nominally due to excess Cs. Silver oxide is ALD deposited from pulses of the organometallic precursor (hfac)Ag(PMe$_3$) and ozone at 200 °C, with a growth thickness rate of 0.04 nm/cycle [25]. As of 2021, cycles can take as few as 0.3 seconds. Atmospheric CVD is a more rapid alternative [26]. Smooth Ag films of ~1nm/minute are formed by PVD techniques [27], [28], followed by an oxide step. Cs vapor must be stochiometrically precise to form the weakly bound semiconductor with as few defects as possible. A precision valved collimated alkali metal Source - AKS- designed for evaporation of elemental high vapor pressure alkali metals like cesium (Cs), is standard in UHV systems. Including the Cs deposition, we anticipate ~30s per nm, about one 10 hour shift per 1µm, realistically 2 µm per 24 hour day. A ~30 µm film could be made in about 3 weeks of 75% operation – 20-30 µm as enough to demonstrate whether this is an interesting material by measuring if it is ohmic or diode contacted; n- or p- type; resistivity as a function of temperature; I-V curves at temperatures; depletion depth; thermal noise; and $V_{breakdown}$. If ALD proves too slow in this project, PVD – particularly plasma enhanced e-beam, and CV .

## *Measurements*

*Electronic Measurements of Completed Diodes:* The carrier type and mobility can be determined before completion of a sealed diode or ion chamber in an Argon glove box with a 4-point "ho"t probe method and by DC Hall mobility $\mu_H$ system. Measurements can determine conductivity $\sigma_o = ne\mu$, the mobility $\mu = R_H/\rho$, carrier concentration $n=(eR_H)^{-1}$, and the sign of the carriers must be made over a range temperatures. Two probes, one heated, will be applied to the materials, and the resulting sign of the induced voltage registered with a picovoltameter or equivalent. I-V curves must be taken at 80°C>T>-40°C. Saturated drift fields of up to 10V/μm should be attempted, and $V_{breakdown}$ found. Noise rates at ~5e-10e levels $Cs_3Sb$ and Ag-O-Cs at room, dry ice, $LN_2$ and LHe temperatures would require a mK-cooled parametric preamp.

*Photoconductive, Electron spot, and Radiation Measurements* Photoconduction can be studied with 4 probes, with light admitted using a quartz fibers covering 275-1300nm, and with a pulsed SEM e-beam across the surface of the disc. Pulse optical measurements should use 10ps wide pulsed lasers. From these measurements as function of distance/time from the light or electron source, limits on the pair energy, and mobility x carrier lifetime/drift length products of the carriers can be inferred by a set of 4 probe waveforms across the semiconductor. A Hall effect rig with 2T magnets can also measure mobility. If the above measurements warrant, using low noise voltage bias sources and charge-sensitive preamps, pulses induced by a 662 KeV cesium, 1.1 MeV Co, 1 MeV beta and an alpha source could measured using a 20bit ADC used for Ge or similar spectroscopy.

*Discussion and Summary*
*Scaling and Costs:* This Pilot program will provide data to decide the merits to follow-on for thicker and larger area ALD detectors and for few cc-volume detectors from melt-grown materials that could vie with Si, Ge, NaI or other detectors. ALD and flood MBE systems are being scaled up to be used in large flat-panel displays and solar energy on ~$2m^2$ scales. Large area films could be patterned or diced into smaller projected areas. Costs of detectors in production could be informed based on experience of this work.

*Radiation Damage and Sampling Calorimetry:* Remarkably, radiation damage in these weakly bound inorganic semiconducting compounds is low. Comparative tests on gamma- and neutron- irradiation of industrial photomultipliers and various PMT window materials confirm that permanent degradation of PMT sensitivity depends almost entirely on PMT window transmission loss and phosphorescence, not on photocathode or dynode degradation [29],[30],[31] with *no obvious effect either on the QE of the photocathode or the gain.*

*Damage Annealing*: The formation temperatures are low (300-350°C) which enables rapid annealing yet without decomposition if kept at lab temperatures (i.e. photocathodes with high current draws resistively heat and decompose). It is posited that in $Cs_3Sb$ dislocated large Cs atoms migrate rapidly back into the Sb matrix; pure silver oxides are as rad-hard as most other metal-oxides. $Cs_3Sb$ or Ag-O-Cs "tiles" interleaved with absorber plates, similar to Si-tile calorimeters, might yield very fast sampling calorimeters in high-dose regions, surviving where Si would not.

*$Cs_3Sb$ tile calorimeters and high rate MIP Detectors:* could operate relatively uncooled, with 4x larger dE/dx, 1.8x e-hole pairs per energy, higher radiation resistance and particularly the ability to operate at much highe rates than Si without polarization effects since the hole and e mobilities are nearly equal and large.

*CsSbPM*: A $Cs_3Sb$ micro-avalanche array detector similar to a SiPM, a "CsSbPM", would operate at 300°K with no cooling.

*Atomic Layer and Ultra-thin Film Techniques - Applications:* A pilot program using atomic layer techniques may prove important for fabricating new collected ion materials, Ohmic or diode electrodes, very low mass low Z gas exclusion/protection (graphene, BN), and new engineered scintillators – as an example, fabrication of perfected ZnO:Ga on tiles (0.1ns/0.7ns rise/decay;largest photons per 1ns). Using graphene or BN films to seal hydroscopic or oxidizing scintillators may be useful for low energy experiments.

A) *Applications Elementary Particle Physics:* semiconductor detectors with low pair-energy thresholds range over detecting dark matter experiments, very low energy (sub-MeV, coherent) ν detection, neutrino-less double-β decay, ν mass, and ν interactions/oscillations from cosmic rays, radiosource/reactor and solar ν's.

B) *Applications Energy/Intensity Frontiers*: $Cs_3Sb$ and Ag-O-Cs materials have #ions/ΔE, dE/dx, μ 2-3x larger than Si, with radiation resistance, possibly enabling very high rate, high resolution sampling calorimeters, dE/dx counters, and tracking.

C) *Atomic layer deposition techniques* offer the potential to make nm/submicron stacks of different materials, including p- and n- dopants for diodes, electrodes, transparent electrodes, hermetic seals, and potentially active materials serving as neutron converters.

D) *Further Applications* include low energy nuclear spectroscopy replacing Ge detectors, x-ray fluorescence detectors, mass spectrometer detectors, biomedicine using isotopes or x-rays, low energy industrial x-ray scanners, night-vision equipment not requiring cooling, and homeland security. Inorganic scintillators (doped ZnO, Cs or Na iodides, others) may benefit from applying atomic layer techniques for protection or fabrication. Radiation-hard sensors that operate at very high rates with possible implementations in medical devices such as Rapid CT or 10-20 ps PET, industrial/cargo 0.5-20 MeV gamma scanning CT, analytical instrumentation (fast X-ray spectroscopy and high speed scanning electron microscope analytics).